\documentclass{article}

\usepackage{amsfonts}
\usepackage{amsmath}
\usepackage{amssymb}

\def\be{\begin{eqnarray}}
\def\ee{\end{eqnarray}}
\def\nn{\nonumber}

\def\p{\partial}

\def\Tr{{\rm Tr}\,}

\newdimen\linethick  \linethick=0.4pt
\newdimen\hboxitspace    \hboxitspace=5pt
\newdimen\vboxitspace    \vboxitspace=5pt

\hoffset= - 1.3in         

\title{{\bf On genus expansion of knot polynomials\\
and hidden structure of Hurwitz tau-functions } \vspace{.2cm}}
\author{{\bf A.Mironov}\footnote{ {\small {\it
Lebedev Physics Institute} and {\it ITEP, Moscow, Russia}};
mironov@itep.ru; mironov@lpi.ru}, {\bf A.Morozov}\thanks{{\small
{\it ITEP, Moscow, Russia}}; morozov@itep.ru}, \  and {\bf
A.Sleptsov}\thanks{{\small {\it ITEP, Moscow, Russia}};
sleptsov@itep.ru}\date{ }}

\begin{document}
 \maketitle

\vspace{-5.5cm}

\begin{center}
\hfill FIAN/TD-07/13\\
\hfill ITEP/TH-11/13\\
\end{center}

\vspace{4.5cm}

\centerline{ABSTRACT}

\bigskip

{\footnotesize In the genus expansion of the HOMFLY polynomials
their representation dependence is naturally captured by
symmetric group characters. This immediately implies that the
Ooguri-Vafa partition function (OVPF) is a Hurwitz tau-function. In
the planar limit involving factorizable special
polynomials, it is actually a trivial exponential tau-function. In fact, in
the double scaling Kashaev limit (the one associated with the
volume conjecture) dominant in the genus expansion are terms
associated with the symmetric representations and with the
integrability preserving  Casimir operators, though we stop one step
from converting this fact into a clear statement about the OVPF
behavior in the vicinity of $q=1$. Instead, we explain that the genus
expansion provides a hierarchical decomposition of the Hurwitz
tau-function, similar to the Takasaki-Takebe expansion of the KP
tau-functions. This analogy can be helpful to develop a substitute
for the universal Grassmannian description in the Hurwitz
tau-functions.}

\vspace{.5cm}

\paragraph{\large Introduction.}

The 't Hooft's genus expansion is one of the main tools
to study the non-perturbative physics of Yang-Mills fields.
These days in order to illustrate a significance of this approach
it is enough to mention the AdS/CFT correspondence \cite{AdS/CFT}.
However, we are still far from possessing reliable methods to
deal with Yang-Mills theories in the confinement phase,
the current goal is to develop them in simpler situations,
either with an extended supersymmetry in various dimensions $d\leq 6$
or with topological invariance in $d=2$ and $d=3$.
Right now the center of theoretical studies is concentrated
around the $3d$ topological Chern-Simons (CS) theory \cite{CS},
where the Wilson line observables are nothing but the knot polynomials
\cite{knopols,WitCS},
and it is very important to understand how the genus
expansion looks like and works in this relatively simple
but still insufficiently understood Yang-Mills model.
In particular, since arbitrary correlation functions
in this theory can be exactly calculated, one can test various
general expectation, including integrability properties
of exact non-perturbative partition functions \cite{UFN3}.

The first step in this direction was done in \cite{MMSle},
where it was shown that the most naive of such partition functions
for the CS theory, the Ooguri-Vafa partition function \cite{OV}
is not a tau-function of the simplest KP/Toda type \cite{UFN3},
but instead belongs to a much less investigated class of
the Hurwitz tau-functions \cite{MMN}.
It is the goal of this letter to provide a brief formulation
of this statement.
The Hurwitz tau-functions are more general than the KP/Toda ones,
and their integrability properties still remain obscure.
Only in the planar limit, where the fully factorizable {\it special}
polynomials \cite{DMMSS} arise, an ordinary (actually, trivial)
free fermion tau-function arises.
We demonstrate that the genus expansion implies a natural
hierarchy of deviations from the KP integrability,
somewhat similar to the Takasaki-Takebe description of
how the KP tau-functions themselves deviate from the
quasiclassical (dispersionless) tau-functions.

We also comment briefly on the double scaling Kashaev limit \cite{volcon},
when representations are large and $q-q^{-1}$ is small.
In fact, the symmetric representations and the KP integrability preserving
terms dominate in this limit, still it is not quite easy to
extract a quantity which can be interpreted as a non-trivial KP
tau-function, responsible for description of the volume conjecture:
this remains a challenging open problem.

\paragraph{\large Definitions.}

The central object in Yang-Mills theory is the Wilson loop average
\be
H_R^{\cal K\subset {\cal M}}(G)
\ \sim\ \left< {\rm Tr}_R\ P\exp \left(\oint_{\cal K}  {\cal A}\right)\right>_{CS}
\label{HRave}
\ee
where ${\cal K}$ is a line (or a set of non-intersecting lines) in the space-time
${\cal M}$ and $G$ is the gauge group.
In the case of Chern-Simons theory, the action is topological \cite{AS,WitCS}
\be
\frac{k}{4\pi}\int_{\cal M} \Tr \Big({\cal A}d{\cal A} + \frac{2}{3}{\cal A}^3\Big)
\ee
and the Wilson average does not change under small deformations of the contour ${\cal K}$
(in drastic contrast with the perimeter and area laws for more general Yang-Mills theories),
thus, it actually depends on its linking class only: the
Wilson average is actually a {\it knot function} (for the exact description of its invariance
properties see \cite{WitCS} and subsequent papers \cite{knotCS}).

For a variety of reasons it is convenient to consider the {\it reduced} knot functions,
which are unity for the unknot, this means that the r.h.s. in (\ref{HRave})
is divided by the same average (in the same representation $R$) with ${\cal K}=unknot$,
and we assume this normalization in this letter.
At the same time sometimes the relevant objects are the {\it unreduced} knot functions,
where the unknot factor is restored, then it matters what it is.
The unknot average is often postulated to be equal to the
quantum dimension $\chi_R^*$ of representation $R$, which is a value of the Schur polynomial
$\chi_R(p)$ at a particular point in the space of the time variables $p_k=p_k^*$.
In particular, a clever way to define the generating function for the knot
functions of a given knot in all representations seems to use the unreduced quantities:
\be
Z^{\cal K}(G) = \sum_R H_R^{\cal K}(G|\bar p)\,\chi_R^*\,\chi_R(\bar p)
\ee
We call this quantity Ooguri-Vafa partition function, since it is intensively used
in \cite{OV}. We denote the time-variables here by $\bar p_k$ to distinguish them from
$p_k$ in the {\it extended} knot functions \cite{DMMSS,MMMkn1},
naturally appearing in study of the braid representations and of the character expansions
(despite we do not use them in this paper, we prefer to avoid a confusion in
comparison with the other papers and keep this by now standard notation).

Of course, the average (\ref{HRave}) is defined for a given gauge group,
say, $G=SU(N)$.
In fact, the parameter $N$ enters the answers only in a peculiar combination
with the coupling constant: $A=q^N$, $q=e^{2\pi i/(k+N)}$ and $q$
is actually a quantization parameter of the quantum group $SU_q(N)$.
This follows from the study of braid representations \cite{TR}
and the character expansion \cite{DMMSS,MMMkn1,MMMkn2} of the knot polynomials:
for example, for $R=\Box$ one can use their direct corollary,
{\it the skein relation} \cite{skein,MMpols} and
complete it with the cabling technique \cite{AnoM} for other representations.
In this way one can actually prove that for the simply connected space-time
${\cal M}=S^3$ or $R^3$ the Wilson averages are actually Laurent polynomials
of $q$ and $A$ (however, for more complicated ${\cal M}$ this property
is no longer preserved).

\paragraph{\large Genus expansion  of knot polynomials  \cite{MMSle}.}

In fact, $A$ is defined as an (exponentiated) 't Hooft's coupling constant.
In the 't Hooft's planar limit ($q=1$, $A$ fixed) the Wilson average for a link
(a multi-trace operator)
decomposes into a product of averages over its particular components.
If the cabling technique (see \cite{AnoM,AnoMMMpath} for a review and modern applications)
is used to represent the colored knot as a link made from $|R|$ copies the same knot
in the fundamental representation, this implies
the factorization of {\it the special polynomials} \cite{DMMSS,IMMMfe,LP,chik}:
\be
H_R^{\cal K}(q=1|A) = \sigma_\Box^{|R|}(A), \ \ \ \ \ \ \ \ \ \ \
\sigma_\Box(A) = H_\Box^{\cal K}(q=1|A)
\ee
This is the starting point of the {\it genus expansion}
(also known as the AMM/EO topological recursion \cite{AMM/EO}),
which for the HOMFLY polynomials looks like \cite{MMSle}:
\be
\boxed{
H_R^{\cal K}(q|A) = \Big(\sigma_{_\Box}(A)\Big)^{|R|}
\exp \left(\sum_{\Delta}
\ z^{|\Delta| + l(\Delta)-2}\
S_\Delta\big(A|z^2\big)\
\varphi_R(\Delta)
\right)
}
\label{geH}
\ee
where the sum goes over all Young diagrams $\Delta$, as usual,
$|\Delta|$ and $l(\Delta)$ denote respectively the numbers of boxes and lines
in $\Delta$,
\be
z = \frac{q^2-q^{-2}}{\sigma^2_{_\Box}(A)},
\ee
$S_\Delta (A|z^2)$ are series in powers of $z^2$,
and the coefficients $\varphi_R(\Delta)$
are the subject of the next paragraph.
At the moment it is important only that \cite{MMN,MMN2}
\be
\varphi_R(\Delta) = 0 \ \ \ \ {\rm  for} \ \ \ \ |\Delta|>|R|
\ee

The coefficients of the $z^2$-series $S_\Delta(A,z^2)$
are polynomials in $A$
(the $z^0$ term in $S_{[1]}$ would be logarithmic,
but we explicitly extracted it as a pre-exponential
power of $\sigma_{_\Box}(A)$).
However, the parameter $z$ contains  $\sigma_{_\Box}(A)$
in the denominator.
In lower terms of the $z$-expansion
they are canceled by the pre-exponential,
but in higher powers a sophisticated cancelation
takes place because of some non-trivial relations
between different $S_\Delta(A,z^2)$.

In fact, the expansion looks somewhat simpler in terms of $z' = q-q^{-1}$
(used in \cite{MMSle}),
however, this variable is not-invariant under the change $q\longrightarrow -q$.
Another "obvious" variable $h=\log q$ \ (i.e. $q=e^h$) also has this drawback.
Another disadvantage of $h$ is that the knot polynomial is not a polynomial in this variable,
but in (\ref{geH}) we actually expand a logarithm of $H_R$,
which is {\it not} a polynomial, thus, this may be not of that importance.
So far we did not observe any essential difference between expansions (\ref{geH})
in all the three variables ($z$, $z'$, $h$), though in this text we use $z$
as the most appealing from theoretical point of view (still examples
in \cite{MMSle} are given in $z'$, since the expressions for $\sigma_\Delta$ are
a little shorter in this variable).

The genus expansion for the superpolynomials \cite{sup} should have the same structure
as (\ref{geH}), but only the first term is conjectured so far \cite{Anton,AnoMMM21},
and the situation with this kind of generalization
remains controversial and very interesting.

\paragraph{\large HOMFLY polynomial as a $W$-transform of the character.
\label{char}}

The most spectacular feature of the genus expansion (\ref{geH})
is that the dependence on the representation $R$ is fully encoded in the
{\it extended} symmetric group characters $\varphi_R(\Delta)$,
which are the eigenvalues of the generalized cut-and-join operators
\cite{MMN,MMN2}
\be
\hat W(\Delta)\, \chi_R\, = \,\varphi_R(\Delta)\,\chi_R
\label{Wchi}
\ee
These operators form an interesting commutative algebra \cite{IK,MMN,MMN2}
and their common eigenvectors are the Schur functions $\chi_R$,
the characters of the universal linear group $GL(\infty)$.

Making use of this fact, one can express the HOMFLY polynomial
as an action of a $W$-evolution operator on the character:
(\ref{geH}) and (\ref{Wchi}) together imply that
\be
\boxed{
H_R^{\cal K}(q|A)\chi_R =
\Big(\sigma_{_\Box}(A)\Big)^{|R|}
\exp\left(\sum_\Delta \beta_\Delta^{\cal K} \hat{W}(\Delta)\right)
 \chi_R
}
\label{chaH}
\ee
where the coefficients $\beta_\Delta^{\cal K}$, yet another set of time-variables depend on the knot and on the $z$-variable:
\be
\beta_\Delta^{\cal K} = z^{|\Delta| + l(\Delta)-2}\
S_\Delta^{\cal K}\big(A|z^2\big)
\ee

\paragraph{\large Ooguri-Vafa partition function as a Hurwitz tau-function.
\label{Hur}}

As a by-product of this representation, (\ref{geH}) and (\ref{chaH}) imply a similar representation for
the Ooguri-Vafa partition function:
\be
Z_{OV}^{\cal K}\{\bar p\} = \sum_R H_R^{\cal K}\,\chi_R^*\,\chi_R\{\bar p\}
= \sum_R\sigma_{_\Box}^{|R|}(A)\, \chi_R^*\,
\exp \left(\sum_{\Delta} z^{|\Delta|+l(\Delta)-2}\,
S_\Delta (A|z^2)\, \hat{ W}(\Delta)\right) \chi_R\{\bar p\} = \nn \\
= \exp \left(\sum_{\Delta} z^{|\Delta|+l(\Delta)-2}\,
S_\Delta^{\cal K} (A|z^2)\, \hat{W}(\Delta)\right)
\exp \left(\sum_k \frac{\sigma_\Box^{\cal K}(A)}{k}p_k^*\bar p_k\right)
=
\boxed{
\exp\left(\sum_\Delta \beta_\Delta^{\cal K} \hat{W}(\Delta)\right)
\tau_0^{\cal K}\{\bar p\}
}
\label{OVW}
\ee
where
\be
\beta_\Delta^{\cal K}(q|A) = z^{|\Delta|+l(\Delta)-1}\, S_\Delta^{\cal K} (A|z^2),\ \ \ \ \ \  \
\tau_0^{\cal K}\{\bar p\}=\exp \left(\sum_k \frac{\sigma_\Box^{\cal K}(A)}{k}p_k^*\bar p_k\right)
\ee
and the cut-and-join operator $\hat W(\Delta)$ acts on the time-variables $\bar p_k$ in this formula.
We used here the celebrated Cauchy formula
\be
\sum_R \mu^{|R|}\chi_R\{p\}\chi_R\{\bar p\} = \exp\left(\sum_k \frac{\mu^kp_k\bar p_k}{k}\right)
\ee

The main implication of (\ref{OVW}) is that
the Ooguri-Vafa partition function is actually a Hurwitz tau-function \cite{MMN}
\be
\tau_H\{\beta|\bar p\} =  \exp\left(\sum_\Delta \beta_\Delta  \hat{\bar W}(\Delta)\right)
\tau_0 \{\bar p\},
\ \ \ \ \ \ \
\tau_0\{\bar p\} = \exp \left(\sum_k c_k\bar p_k\right)
\ee
taken at a particular knot-dependent value of the Hurwitz time-variables $\beta_\Delta$.

\paragraph{\large Linear vs non-linear evolution.}

Any system of commuting operators, like $\{\hat W(\Delta)\}$ can be
considered as a system of Hamiltonians of an integrable system,
and matrix elements of the evolution operator
\be
\hat{\cal U}\{\beta\} = \exp\left(\sum_\Delta \beta_\Delta \hat W(\Delta)\right)
\ee
generate an object deserving the name of tau-function.

However, there is a question of whether the Hamiltonians are independent.
As well known in the theory of renormalization group \cite{MMreng},
it is important to distinguish between {\it linearly} and
{\it algebraically} independent generators:
the best possible example is provided by the multi- and single-trace
operators in matrix models.
In ordinary integrable systems evolutions are always generated
by algebraically independent (single-trace) operators,
while their non-linear combinations (multi-trace operators)
are not included into the definition of the tau-function.

However, an exact relation of the evolution generated by linearly
independent operators and that generated by the algebraically
independent operators, remains an important open question in the
theory of integrable hierarchies, closely related (but not equivalent)
to the problem of {\it equivalent hierarchies} \cite{Kharequiv}.

In our case, the $\hat W(\Delta)$-operators are all linearly independent,
while they all being algebraic functions of a smaller set of the Casimir operators.
According to general principles \cite{GKM2,AMMN}, the Casimir evolution possesses ordinary
(KP/Toda) integrability properties, however, the way the generic
(non-KP) $W$-evolution is expressed through it, is unknown.
Still, it is this, more general $W$-evolution, which the
knot polynomials are related to.
At the same time, there is an important hierarchical parameter
$z^{l(\Delta)}$ in (\ref{chaH}), which measures the algebraic
complexity of $\hat W(\Delta)$, and this provides an
interesting hierarchy of deviations from the KP integrability.
It is subject of the remaining paragraphs of this letter.

\paragraph{\large Hurwitz tau-function via Casimir operators.}

For arbitrary values of $\beta_\Delta$
the Hurwitz tau-function is not an ordinary KP/Toda tau-function
in variables $\bar p_k$. These latter are generated by the Casimir operators $\hat{C}(k)$:
\be
\tau_{KP}\{t|\bar p\}
=  \exp\left(\sum_k t_k \hat{\bar C}(k)\right)
\tau_0 \{\bar p\}
\ee
Like $\hat W(\Delta)$, the Casimir operators have the Schur functions $\chi_R\{\bar p\}$
as their common eigenfunctions,
\be
\hat C(k) \,\chi_R = c_R(k)\chi_R
\ee
with the eigenvalues for the representation given by the Young diagram $R = \{r_1\geq r_2\geq \ldots\geq 0\}$ being
\be
c_R(k) = \sum_{j=1}^{l(R)} \Big((r_j-j+1/2)^k - (-j+1/2)^k\Big)
\label{evC}
\ee
The shift $1/2$ can be substituted by any other one, this induces a linear
transformation of the set of the Casimir operators, but the particular choice of
$1/2$ is more convenient for many purposes, including application to the genus expansion:
see a comment after eq.(\ref{OVC}).

The cut-and-join operators are non-linear combinations of the Casimir operators:
\be
\hat{\bar W}(\Delta) = \sum_{Q} w^\Delta_{Q}\, \hat C(Q)
\ee
where for the Young diagram $Q=\{q_1\geq q_2\geq\ldots\geq 0\}$
\be
\hat C(Q) = \prod_{j=1}^{l(Q)} \hat C(q_i), \ \ \ \ \ \
\frac{\p}{\p t_Q} = \prod_{j=1}^{l(Q)}\frac{\p}{\p t_{q_i}}
\ee
This means that
\be
\boxed{
\tau_H\{\beta|\bar p\} = \left.
\exp\left(\sum_{\Delta,Q} \beta_\Delta
 w^\Delta_{Q}\ \frac{\p}{\p t_Q}
\right)
\tau_{KP}\{t|\bar p\}\right|_{t=0}
}
\ee
The action of this operator, namely, the terms  with $l(Q)>1$,
break the KP/Toda integrability.
Hopefully, the generic Hurwitz tau-function belongs to the class
of generalized tau-functions of \cite{GKLMM}, but this is yet
an open question.

\paragraph{\large Genus expansion via Casimir operators.}

Instead of (\ref{OVW}) one can express the genus expansion formulas of
\cite{MMSle} through the Casimir operators:
\be
\boxed{
Z_{OV}^{\cal K}\{\bar p\} =
\exp \left(\sum_{\Delta} z^{|\Delta|+l(\Delta)-2}\,
\tilde S_\Delta^{\cal K} (A|z^2)\, \hat{C}(\Delta)\right)
\tau_0^{\cal K}\{\bar p\}
}
\label{OVC}
\ee
Here $\tilde S_\Delta$ are new combinations of the higher special polynomials
$S_\Delta$, and if the shift in (\ref{evC}) is chosen to be $1/2$,
they are also series in {\it even} powers of $z$.

\paragraph{\large Large-$R$ behavior.}

Since for representations of large sizes  $|R|$ the eigenvalues of the Casimir
operators grow as
\be
C_R(k) \sim \gamma_k|R|^k
\ee
it is clear that the growth of the symmetric group characters is bounded by
\be
\varphi_R(\Delta) \lesssim  |R|^{|\Delta|} \ \ \ \ \
\ee
This means that in the large $R$ limit eq.(\ref{OVC}) implies for the HOMFLY polynomial
at the generic value of $A$:
\be
\log H_R = |R|\!\!\!\! \sum_{\Delta:\ \ l(\Delta)=1}
(z|R|)^{|\Delta|}\Big( \gamma_{|\Delta|}\sigma_\Delta(A|0) + O(z)\Big)
\ee
which means that in the double scaling limit
(used, for example, in the context of the volume conjecture \cite{volcon})
\be
z\longrightarrow 0,\ \ \ \ \ \ \ \  |R|\longrightarrow \infty, \ \ \ \ \ \ \ \ \
u=z|R| \ \ \ {\rm fixed}
\label{vclim}
\ee
the dominant contribution to $\log H_R$ proportional to $|R|$ is provided by
the sum over the symmetric representations $\Delta$ with single row
Young diagrams, $l(\Delta)=1$.

Ironically, this does not have {\it direct} implications for
the volume conjecture {\it per se}: the thing is that it is formulated
in the case when $N$ is fixed rather than $A$, so that also $A=q^N\longrightarrow 1$,
and $\sigma_\Delta(A=1|0)=0$ for the symmetric representations
(this property is implied, for example, by the reduction property \cite{DMMSS}
${\aleph}_R(q) = {\aleph}_\Box(q^{|R|})$ of the Alexander polynomials
for the single hook Young diagrams $R$, see \cite{MMSle}).
Therefore, the situation with the volume conjecture in the context of the
genus expansion is more tricky \cite{MMSle,MMpols}.
This story has a lot to do
with the Mellin-Morton-Rozansky expansion \cite{MeMoRo} into the inverse Alexander
polynomials.

Another obvious observation is that at the same limit (\ref{vclim})
dominant in (\ref{OVC}) are the terms with $l(\Delta)=1$, i.e. {\it linear}
in the Casimir operators. If all other terms were simply thrown away,
the Hurwitz tau-function would reduce to a KP one, i.e. we would have
a naive KP/Toda integrability for the Ooguri-Vafa partition function.
Unfortunately, things are again not so simple:  $Z_{OV}$ is defined as
a sum over representations $R$, so that $R$ is not a free parameter, which
one can adjust in a desired way.

One can put it differently, formulating the claim in terms of the Pl\"ucker relations: it is well-known that
in order for a linear combination of
the Schur functions $\chi_R$ to be a KP tau-function, the coefficients of this combinations have
to satisfy the Pl\"ucker relations \cite{Pl}. In the case under consideration this property (the Pl\"ucker relations)
is satisfied only asymptotically at large $|R|$.

\paragraph{\large Genus expansion for knot polynomials vs Takasaki-Takebe expansion.}

Despite an exact reduction to the KP integrability fails, the
appearance of the $z$-variable in (\ref{OVC}) remains very suggestive.
In particular, it resembles the famous Takasaki-Takebe
description \cite{TaTa} of quasiclassical expansion for the KP tau-functions
around their dispersionless approximations.
That is, they demonstrate that the quasiclassical limit is described
by a nearly-diagonal matrices in the universal Grassmannian,
while contributions from every next sub-diagonal is damped by
an extra power of $\hbar$.
In the free fermion representation of the KP $tau$-functions
this is expressed as follows:
\be
\tau_{KP}\{u_L(z)|t_k\} =
\left<\exp\left(\sum_k t_k H_k\right)\exp \left(\frac{1}{\hbar}
\sum_{L} \oint dzu_L(z)\, \bar\psi (z) (\hbar \p_z)^L \psi (z)\right)\right>
\label{TaTa}
\ee
where $u_1(z)$ parameterizes the quasiclassical
(dispersionless) tau-functions, and further terms of the loop $\hbar$-expansion are associated with $u_L(z)$, which,
in their turn, are associated with the $W^{(L+1)}$ algebra. More exactly, the terms
$\oint dzz^{n+L}\, \bar\psi (z) (\hbar \p_z)^L \psi (z)$ correspond to the action of $W^{(L+1)}_n$-generators on
the tau-function.

In (\ref{TaTa}) $H_k=\sum\psi_l\psi^*_{l+k}$ are the Hamiltonians in the free fermion representation giving rise to the
KP flows, and the average is defined w.r.t. fermionic vacuum, see \cite{TaTa,versus} for notation and details.
Thus, the integral in (\ref{TaTa}) gives a specific parameterization of the group element
\be
g = :\exp \left(\oint\oint U(z,z')\bar\psi(z')\psi(z)\right):
\ee
parameterizing the generic KP or Toda-lattice tau-function,
\be
\tau_{Toda}\{\bar t, t |g\} = \left<\exp\left(\sum_k \bar t_k \bar H_k\right)\ g\
\exp\left(\sum_k t_k H_k\right)\right>
\ee
and consistent with the quasiclassical expansion and formula (\ref{TaTa}) describes how the KP tau-function is formed from the
quasiclassical one, as a series in $\hbar$.

\bigskip

Actually, eq.(\ref{OVC}) describes in a very similar way how the Hurwitz tau-function
is formed from the KP one as a series in $z$.
Instead of the sub-diagonal terms in the universal Grassmannian,
the higher corrections in $z$ are associated with the higher powers
of Casimir operators.
This analogy made possible by introduction of the auxiliary parameter $z$
in the generic Hurwitz tau-function in the way inspired by the natural problem of
the genus expansion of knot polynomials,
can shed some light on what is a substitute of the universal Grassmannian \cite{UniGr}
as a universal moduli space \cite{UMS} of the Hurwitz tau-functions.
This can also help to develop a substitute of the free fermion representation
and embed Hurwitz functions into the general (so far badly studied)
world of generalized tau-functions of \cite{GKLMM},
associated with arbitrary Lie algebras.

\paragraph{\large Acknowledgements.}

Our work is partly supported by Ministry of Education and Science of
the Russian Federation under contract 8207, the Brazil National Counsel of Scientific and
Technological Development (A.Mor.), by NSh-3349.2012.2,
by RFBR grants 13-02-00457 (A.Mir. and A.S.) and 13-02-00478 (A.Mor.),
by joint grants 12-02-92108-Yaf, 13-02-90459-Ukr-f, 13-02-91371-ST, 14-01-93004-Viet.

\end{document}